\documentclass[letterpaper,final,12pt]{iopart}

\expandafter\let\csname equation*\endcsname\relax
\expandafter\let\csname endequation*\endcsname\relax

\usepackage{amsmath}

\usepackage{graphicx}
\usepackage[colorlinks=true,linkcolor=blue,citecolor=blue,urlcolor=black,breaklinks=true]{hyperref}%

\usepackage{iopams}  

\usepackage[numbers, square, sort&compress]{natbib}
\def\onlinecite{\citenum}

\def\apss{Astrophys. Space Sci.}
\def\apsst{Astrophys. Space Sci. Trans.}

\def\solphys{Sol. Phys.}

\def\ssr{Space Sci. Rev.}

\def\grl{Geophys. Res. Lett.}

\def\jgr{J. Geophys. Res.}
\def\jgra{J. Geophys. Res. (Space Physics)}

\def\physscr{Phys. Scr.}

\def\epjd{Eur. Phys. J. D}
\def\phfl{Phys. Fluids}
\def\phflb{Phys. Fluids B}
\def\phpl{Phys. Plasmas}
\def\ppcf{Plasma Phys. Contr. Fusion}
\def\jplph{J. Plasma Phys.}

\def\jpsj{J. Phys. Soc. Japan}

\def\jpha{J. Phys. A}

\def\adspr{Adv. Space Res.}
\def\npgeo{Nonlin. Proc. Geophys.}
\def\epl{Europhys. Lett.}

\begin{document}

\title[Electron beam-plasma interaction and suprathermal electrons]{Electron beam-plasma interaction and electron-acoustic solitary waves in a plasma with suprathermal electrons}

\author[A~Danehkar]{A~Danehkar$^{1,2}$\footnote{Present address: Harvard-Smithsonian Center for Astrophysics, 60 Garden Street, Cambridge, MA 02138, USA}}

\address{$^{1}$ Centre for Plasma Physics, Queen's University Belfast, Belfast BT7~1NN, UK}
\address{$^{2}$ Department of Physics and Astronomy, Macquarie University, Sydney, NSW 2109, Australia}
\ead{ashkbiz.danehkar@cfa.harvard.edu}
\vspace{10pt}
\begin{indented}
\item[]Received 30 December 2017, revised 26 March 2018\\ 
Accepted for publication 6 April 2018
\end{indented}

\begin{abstract}
Suprathermal electrons and inertial drifting electrons, so called electron beam,  are crucial to the nonlinear dynamics of electrostatic solitary waves observed in several astrophysical plasmas. In this paper, the propagation of electron-acoustic solitary waves is investigated in a collisionless, unmagnetized plasma consisting of cool inertial background electrons, hot suprathermal electrons (modeled by a $\kappa$-type distribution), and stationary ions. The plasma is penetrated by a cool electron beam component. A linear dispersion relation is derived to describe small-amplitude wave structures that shows a weak dependence of the phase speed on the electron beam velocity and density. A (Sagdeev-type) pseudopotential approach is employed to obtain the existence domain of large-amplitude solitary waves, and investigate how their nonlinear structures depend on the kinematic and physical properties of the electron beam and the suprathermality (described by $\kappa$) of the hot electrons. The results indicate that the electron beam can largely alter the electron-acoustic solitary waves, but can only produce negative polarity solitary waves in this model. While the electron beam co-propagates with the solitary waves, the soliton existence domain (Mach number range) becomes narrower (nearly down to nil) with increasing the beam speed and the beam-to-hot electron temperature ratio, and decreasing the beam-to-cool electron density ratio in high suprathermality (low $\kappa$). It is found that the electric potential amplitude largely declines with increasing the beam speed and the beam-to-cool electron density ratio for co-propagating solitary waves, but is slightly decreased by raising the beam-to-hot electron temperature ratio. 
\end{abstract}

%
%
\submitto{\PPCF}
%
\maketitle
%
%

\section{Introduction \label{introduction}}

Nonlinear electron-acoustic solitary waves (EAWs)
are typically produced in a plasma composed of two electron temperature
populations, so called \textit{cool} and \textit{hot} electrons
\citep{Watanabe1977,Ikezawa1981,Mace1990,Hellberg2000}, which have been observed in many space
plasmas \citep{Dubouloz1991,Feldman1983,Bale1998,Pottelette1999,Cattell2002,Baluku2011}. 
In
such a plasma, the wave restoring force is provided by the thermal pressure of
the hot electrons, whereas the cool electrons provide the inertia to support
electron-acoustic oscillations. 
The phase speed of the
EAWs is typically intermediate between the cool and hot electron thermal velocities. 
It has been pointed out that the EAWs only
propagate within a restricted range of the physical conditions, i.e., the hot-to-cool
electron number density ratio between 1/4 and 4, and the hot-to-cool 
electron temperature ratio is greater than 10 
\citep{Gary1985,Mace1990,Mace1999}. As the EAW frequency is much higher than
the ion plasma frequency, the ions may not largely affect the EAWs, and they usually play the
role of a neutralizing background in the plasma.

The observations of the upper layer of the Earth's magnetosphere revealed the injection of
magnetic field-aligned drifting electrons from solar winds, the so-called \textit{electron beam}, into
two different (cool and hot) electron populations 
\citep{Tokar1984,Lin1984,Ogilvie1984,Pottelette1990,Dubouloz1991a,Matsumoto1994,Cattell1998,Tsurutani1998}, which
may excite the EAWs and modify their existence conditions and nonlinear wave structures. It has been found
that the auroral broadband electrostatic noise (BEN) emissions observed in the
Earth's magnetosphere strongly correlate with the magnetic field-aligned electron beams 
\citep{Dubouloz1991a}. In particular, the high time-resolution GEOTAIL
observations of the Earth's auroral region showed that the BEN consists a series of nonlinear localized
electrostatic solitary waves, which are associated with the nonlinear dynamics of
the electron beam instability \citep{Matsumoto1994}. The generation of electrostatic solitary waves was also reported in the polar cap boundary layer (PCBL) region, which are locally involved electron beams \citep{Tsurutani1998}. More recently, the Magnetospheric Multiscale (MMS) Mission observations of the Earth's magnetosphere revealed electrostatic solitary waves in the magnetosheath and magnetopause, likely supported by a \textit{cool} ($\sim 1$--20\,eV) field-aligned drifting  electron component \citep{Ergun2016,Holmes2018}. The effects of 
electron beams have been investigated in EAWs
\citep{Berthomier2000,Mace2001,Sahu2004,ElTaibany2005a,ElTaibany2005b,ElLabany2005,Elwakil2007,Devanandhan2011b,Singh2011,Lakhina2008a,Lakhina2011}. 
Electron beams allow the propagation of EAWs with velocity related to the
beam speed, whereas their amplitude and width depend on the beam physical properties
\citep{Berthomier2000,Mace2001}. It has been found that a hole (hump) in the cool (hot) electron number density, which is not present without electron beam, allows the propagation of positively polarized electrostatic solitary waves \citep{Berthomier2000,Sahu2004,ElTaibany2005a,ElTaibany2005b}.

A population of energetic \textit{hot} ($\sim$ 1--20 keV at Earth's bow shock \cite{Gosling1989}) electrons with \textit{suprathermal} distributions has been
reported in various space plasmas \citep{Pierrard2010}. This population was
found to have a suprathermal (or non-thermal) tail on its velocity distribution
function, and its kinetic energy is much higher than the thermal energy of the
background cool inertial electrons. It has been shown that these suprathermal hot electrons
are well described by a family of $\kappa$-distribution functions
\citep{Vasyliunas1968}. The deviation from a Maxwellian distribution is
described by the spectral index $\kappa$, i.e., low values of $\kappa$ are
associated with a significant suprathermality, whereas a Maxwellian
distribution is recovered in the limit $\kappa\rightarrow\infty$
\citep{Summers1991,Baluku2008,Hellberg2009}. Suprathermal (and non-thermal) electrons have recently been incorporated into theoretical models of EAWs
\citep{Sahu2010,Younsi2010,Danehkar2011,Devanandhan2011a,Sultana2012,Sultana2012b,Sahu2013,Han2014b,Han2014c,Singh2016}. 
It has been found that higher deviations from a Maxwellian raise negative
polarity electron-acoustic solitary waves
\citep{Sahu2010,Younsi2010,Danehkar2011,Devanandhan2011a}, though the soliton
existence regions become narrower \citep{Younsi2010,Danehkar2011}. Moreover,
electron-acoustic shock structures were found to become narrower and steeper
in higher suprathermality \citep{Sultana2012}. 

Recently, a number of papers have been devoted to the effects of both electron beams and suprathermal
electrons on EAWs \citep{Singh2011,Devanandhan2011b,Danehkar2011a,Danehkar2014,Singh2016}. 
However, the positive polarity signature,
which was previously found for EAWs \citep{Berthomier2000}, was not reproduced
using $\kappa$-distributed electrons \citep{Devanandhan2011b}, which might be
due to the chosen parameter range. Previously, a plasma with suprathermal
electrons, inertial warm electron beam, inertial cool electron and
inertial (mobile) ions were found to produce both positive and negative electrostatic
potentials \citep{Singh2011}. To have positively polarized electrostatic waves, the
inertia of the warm electrons, not the beam velocity, seems to be
important \citep{Singh2011}. However, it requires a rather high density of cool
electrons \citep{Lakhina2008a,Lakhina2011,Singh2011}, which does not seem to
correspond to available observations \citep{Verheest2005,Verheest2007}. 
Alternatively,  mobile ions \citep{Danehkar2009,Saini2010,Saberian2013} or inertial positrons (or electron holes) \citep{Jilani2014,Danehkar2017} can provide the inertia for the propagation of
positive polarity electrostatic waves with suprathermal electrons in the slow (ion) or fast (positron) acoustic modes, respectively. 

In this paper, we aim to investigate the effects of an electron beam component,
together with suprathermal $\kappa$-distributed electrons, on the existence
conditions and properties of the EAWs. A theoretical two-fluid model is presented
in Section~\ref{model}. In Section \ref{DR}, we derive a linear dispersion
relation for the small-amplitude EAWs. In Section \ref{nonlinear}, the Sagdeev pseudopotential method is employed
to obtain the nonlinear solution of the large-amplitude EAWs. The occurrence of electron-acoustic solitary waves is investigated in Section \ref{existence}.
In Section \ref{investigation}, we investigate the electron beam effects on
the characteristics of EAWs. Our conclusions are given in Section \ref{conclusion}.

\section{Theoretical two-fluid model\label{model}}

We consider a four-component collisionless, unmagnetized plasma consisting of cool inertial background electrons (at
temperature $T_{c}\neq0$), cool inertial electron beam (at temperature $T_{b}\neq0$), hot inertialess suprathermal 
electrons modeled by a $\kappa$-distribution (at temperature $T_{h}\gg
T_{b},T_{c}$), and uniformly distributed stationary ions.

The fluid model is governed by the following one-dimensional equations:
\begin{align}
&  \frac{\partial n_{j}}{\partial t}+\frac{\partial(n_{j}u_{j})}{\partial
x}=0,\label{eq_1}\\
&  \frac{\partial u_{j}}{\partial t}+u_{j}\frac{\partial u_{j}}{\partial
x}=\frac{e}{m_{e}}\frac{\partial\phi}{\partial x}-\frac{1}{m_{e}n_{j}}%
\frac{\partial p_{j}}{\partial x},\label{eq_2}\\
&  \frac{\partial p_{j}}{\partial t}+u_{j}\frac{\partial p_{j}}{\partial
x}+\gamma p_{j}\frac{\partial u_{j}}{\partial x}=0,\label{eq_3}
\end{align}
\begin{align}
n_{h}(\phi)  &  =n_{h,0}\left[  1-\frac{e\phi}{k_{B}T_{h}(\kappa-\tfrac{3}%
{2})}\right]  ^{-\kappa+1/2}\,,\label{eq_4}\\
Zn_{i,0}  &  =n_{c,0}+n_{b,0}+n_{h,0},\label{eq_5}\\
\frac{\partial^{2}\phi}{\partial x^{2}}  &  =-\frac{e}{\varepsilon_{0}}\left(
Zn_{i,0}-n_{c}-n_{b}-n_{h}\right)  , \label{eq_6}%
\end{align}
where $n_{j}$, $u_{j}$ and $p_{j}$ are, respectively, the number density, the
velocity and the pressure of the cool background electrons (denoted by index $j=c$) and
the electron beam (denoted by index $j=b$), $n_{h,0}$ and $T_{h}$ are,
respectively, the equilibrium number density and the temperature of the hot
electrons modeled by the $\kappa$-distribution equation (\ref{eq_4})
\citep{Summers1991,Baluku2008,Hellberg2009}, the spectral index $\kappa$
measures the deviation from a Maxwellian distribution ($\kappa>3/2$),
$n_{i,0}$ the undisturbed ion density, $Z$ the number of ions, $\phi$ the
electrostatic wave potential, $e$ the elementary charge, $m_{e}$ the electron
mass, $\varepsilon_{0}$ the permittivity constant, $k_{B}$ the Boltzmann
constant, and $\gamma=(f+2)/f$ denotes the specific heat ratio for $f$ degrees
of freedom. For the adiabatic (cool and beam) electrons in 1-D ($f=1$), we get $\gamma
=3$. The ions are assumed to be stationary, i.e., $n_{i}=n_{i,0}=$ const. at
all times. As the plasma is quasi-neutral at equilibrium, we have Eq.
(\ref{eq_5}), where $n_{c,0}$ and $n_{b,0}$ are the equilibrium number density of the cool background electrons and the electron beam, respectively. 
All four components are coupled via the Poisson's equation
(\ref{eq_6}).

Normalizing equations (\ref{eq_1})--(\ref{eq_6}) by appropriate quantities, we
obtain a dimensionless set of fluid equations as follows:%
\begin{align}
&  \frac{\partial n_{j}}{\partial t}+\frac{\partial(n_{j}u_{j})}{\partial
x}=0,\label{eq_11}\\
&  \frac{\partial u_{j}}{\partial t}+u_{j}\frac{\partial u_{j}}{\partial
x}=\frac{\partial\phi}{\partial x}-\frac{\theta_{j,h}}{n_{j}}\frac{\partial
p_{j}}{\partial x},\label{eq_12}\\
&  \frac{\partial p_{j}}{\partial t}+u_{j}\frac{\partial p_{j}}{\partial
x}+3p_{j}\frac{\partial u_{j}}{\partial x}=0,\label{eq_13}\\
\frac{\partial^{2}\phi}{\partial x^{2}}  &  =-\left(  1+\rho_{h,c}+\rho
_{b,c}\right)  +n_{c}\nonumber\\
&  +\rho_{b,c}n_{b}+\rho_{h,c}\left(  1-\frac{\phi}{\kappa-\tfrac{3}{2}%
}\right)  ^{-\kappa+1/2}, \label{eq_17}%
\end{align}
where the density variables $n_{c}$ and $n_{b}$ are, respectively, normalized
by $n_{c,0}$ and $n_{b,0}$, the velocities $u_{c}$ and $u_{b}$ scaled by the hot electron
thermal speed $c_{th}=\left(  k_{B}T_{h}/m_{e}\right)^{1/2}$, the wave
potential $\phi$ scaled by $k_{B}T_{h}/e$, and time and space scaled by the plasma
period $\omega_{pc}^{-1}=\left(  n_{c,0}e^{2}/\varepsilon_{0}m_{e}\right)
^{-1/2}$ and the characteristic length $\lambda_{0}=\left(  \varepsilon
_{0}k_{B}T_{h}/n_{c,0}e^{2}\right)  ^{1/2}$, respectively. The normalized beam speed is defined as $V_{b}=u_{b,0}/c_{th}$, where $u_{b,0}$ is the beam velocity at equilibrium. 
The hot-to-cool
electron number density ratio, $\rho_{h,c}$, and the beam-to-cool electron
number density ratio, $\rho_{b,c}$, are respectively defined as%
\[%
\begin{array}
[c]{cc}%
\rho_{h,c}=n_{h,0}/n_{c,0}, & \text{ \ \ }\rho_{b,c}=n_{b,0}/n_{c,0},
\end{array}
\]
which imply $Z{n_{i,0}}/{n_{c,0}}=1+\rho_{h,c}+\rho_{b,c}$. 

Landau damping is minimized if $0.2\lesssim n_{c,0}/(n_{c,0}+n_{h,0})\lesssim0.8$
\citep{Gary1985,Mace1990,Mace1999}. We consider the region $0.25\leqslant\rho_{h,c}\leqslant4$ in which the linear waves are not strongly damped. The
cool-to-hot electron temperature ratio, $\theta_{c,h}$, and the
beam-to-hot electron temperature ratio, $\theta_{b,h}$, are
defined respectively as
\[%
\begin{array}
[c]{cc}%
\theta_{c,h}=T_{c}/T_{h}, & \text{ \ \ }\theta_{b,h}=T_{b}/T_{h}.
\end{array}
\]
The linear waves survive Landau damping if $T_{h}/T_{c}\gg 10$
\citep{Gary1985,Mace1990,Mace1999}, so we consider the region $\theta_{c,h}\ll0.1$ where the waves are not strongly damped. 
Similarly, electron-acoustic solitary waves excited by the electron beam are weakly damped when the hot to beam electron temperature 
ratio becomes more than 10. We notice that the observations of the BEN in the auroral zone indicated that the presence of magnetic field-aligned electron beams with $T_{b} \sim 1$\,eV, while hot electron population with $T_{h} \sim 500$\,eV \citep{Schriver1989}, or $T_{b} \sim 50$\,eV and $T_{h} \sim 200$--990\,eV \citep{Dubouloz1991a}. Moreover, the observations of the PCBL region pointed $T_{b} \sim 100$\,eV and $T_{h} \sim 25$\,keV \citep{Tsurutani1998}. Recently, the MMS observations of the magnetosphere plasma reported field-aligned drifting electrons with $T_{b} \sim 1$--20\,eV \citep{Ergun2016,Holmes2018} along with hot electrons having $T_{h} \sim 1$\,keV \citep{Ergun2016}. Hence, we can assume $\theta_{b,h}\ll0.1$.

To preserve the excitation of electrostatic waves, there should be no wave magnetic effects and no current from Amp\`{e}re's law \citep{Verheest2009a}, i.e.  $j_{b,0}=n_{b,0}u_{b,0}\simeq 0$. Thus, the electron beam speed $V_{b}$ and the
beam-to-cool electron number density ratio $\rho_{b,c}$ should satisfy the trivially current condition $\rho_{b,c} V_{b} \ll 1 $ \citep{Saini2010}, in addition to the weakly damped condition $\theta_{b,h}\ll0.1$ for the beam-to-hot electron temperature ratio, in order to maintain the excitation of EAWs as the electron beam penetrates into the plasma.

\section{Linear dispersion relation\label{DR}}

To study linear small-amplitude wave solutions of the fluid model, we obtain linearized forms
of equations (\ref{eq_11}) to (\ref{eq_17}). 
We apply a small deviation from the equilibrium state, which produces the derivatives of the first order amplitudes, and keep the expansion up to first order. 
The linear dispersion relation for EAWs is then as follows:%
\begin{align}
1+\frac{\rho_{h,c}}{k^{2}}\left(  \dfrac{\kappa-\frac{1}{2}}{\kappa-\tfrac
{3}{2}}\right)   = & \frac{1}{\omega^{2}-3\theta_{c,h}k^{2}}\nonumber\\
&  +\frac{\rho_{b,c}}{\left(  \omega-kV_{b}\right)  ^{2}-3\theta_{b,h}k^{2}}
\label{eq_23}%
\end{align}
where $\sqrt{3\theta_{c,h}}$ and$\sqrt{3\theta_{b,h}}$ are basically the (normalized)
thermal velocities of the cool electrons and the electron beam, respectively.

Eq. (\ref{eq_23}) indicates that the phase speed ($\omega/k$) increases with raising the 
cool-to-hot electron temperature ratio $\theta_{c,h}=T_{c}/T_{h}$, which is in agreement with 
Ref.~\onlinecite{Danehkar2011}. Although the frequency $\omega(k)$ (also the phase
speed) increases with raising  the beam-to-hot electron temperature ratio $\theta
_{b,h}=T_{b}/T_{h}$, this linear effect is negligible due to small values of
the beam-to-cool electron number density ratio $\rho_{b,c}$. In the limit
$\rho_{b,c}\rightarrow0$ (in the absence of the electron beam),
Eq.~(\ref{eq_23}) recovers precisely Eq.~(14) in 
Ref.~\onlinecite{Danehkar2011}. 

\begin{figure}
[ptb]
\begin{center}
\includegraphics[width=2.9in]{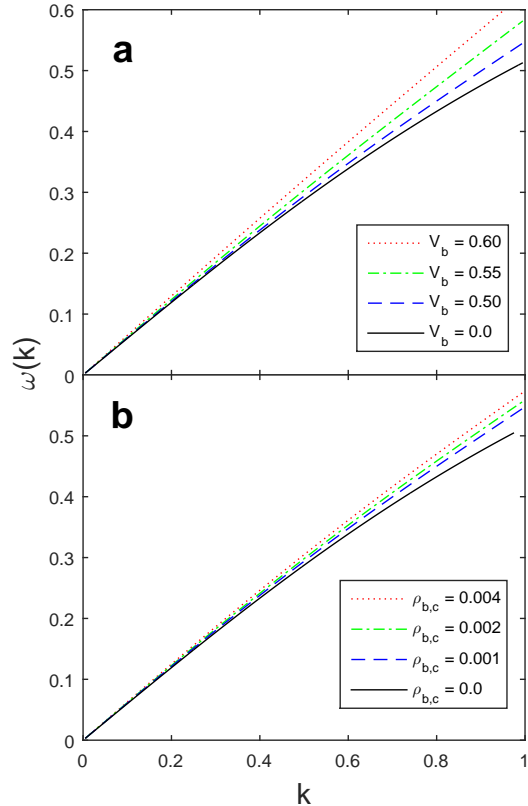}%
\caption{
(a) Variation of the dispersion curve for different values of  
the normalized beam speed $V_{b}$ is depicted. Curves from bottom to top:
$V_{b}=0$ (solid), $0.5$ (dashed), $0.55$ (dot-dashed), and $0.6$ (dotted
curve). Here, $\rho_{b,c}=0.001$. (b) Variation of the dispersion
curve for different values of the beam-to-cool electron number density ratio 
$\rho_{b,c}$. Curves from bottom to top:
$\rho_{b,c}=0.0$ (solid), $0.001$ (dashed), $0.002$ (dot-dashed curve), and
$0.004$ (dotted curve). Here, $V_{b}=0.5$. For both panels, we have taken:
$\rho_{h,c}=1$, $\kappa=3$, and $\theta_{c,h}=\theta_{b,h}=0$.}%
\label{fig1}%
\end{center}
\end{figure}

Figure \ref{fig1} shows the dispersion curve (\ref{eq_23}) of the
electron-acoustic mode in the cold limit ($T_{c}=T_{b}{=0}$). It can be seen
how varying the electron beam parameters $V_{b}$ and $\rho_{b,c}$ affect
the dispersion curve. The phase speed increases weakly with an
increase in $V_{b}$ and $\rho_{b,c}$. Additionally, an increase in the number
density of suprathermal hot electrons or/and the suprathermality (decreasing
$\kappa$) also decreases the phase speed (see Figure 1 in Ref.~\onlinecite{Danehkar2011}).

Following Ref.~\onlinecite{Mace1999}, the wave avoids being strongly damped if its phase velocity 
is between the cool electron and hot electron thermal speed $c_{tc} \ll \omega/k \ll c_{th}$ (or the normalized region here $\sqrt{\theta_{c,h}} \ll \omega/k \ll 1$). Moreover, the wave is weakly damped for $0.2\lesssim k\lambda_{Dc} \lesssim 0.6$, so this region could be of interest only. Here, $c_{tc}=\left(  k_{B}T_{c}/m_{e}\right)^{1/2}$ is the cool electron thermal speed, and $\lambda_{Dc}=\left(  \varepsilon
_{0}k_{B}T_{c}/n_{c,0}e^{2}\right)^{1/2}$ is the cool electron Debye length. As seen in Fig.\,\ref{fig1}, the phase speed of the linear wave is inside the weakly damped region ($0<\omega/k<1$, while $T_{c}=T_{b}=0$). The physical conditions for minimizing weakly Landau damping ($0.25\leqslant\rho_{h,c}\leqslant4$, $\theta_{c,h}\ll0.1$) may also prevent the phase velocity of the waves being strongly Landau damped by either of the cool or hot electrons.

\section{Nonlinear pseudopotential approach\label{nonlinear}}

To obtain nonlinear large-amplitude wave solutions of the fluid model, it is convenient to consider a stationary frame moving with a constant normalized velocity $M$, so called the Mach number. This implies the transformation $\xi=x-Mt$ that replaces the space and time derivatives with $\partial/\partial x=d/d\xi$ and $\partial/\partial t=-Md/d\xi$, respectively. 
Eqs. (\ref{eq_11})-(\ref{eq_17}) in the corresponding reference frame take the following form:
\begin{align}
&  -M\dfrac{dn_{c}}{d\xi}+\frac{d(n_{c}u_{c})}{d\xi}=0,\label{eq_26} \\
&  -M\dfrac{du_{c}}{d\xi}+u_{c}\dfrac{du_{c}}{d\xi}=\dfrac{d\phi}{d\xi}-\frac{\theta_{c,h}}%
{n_{c}}\dfrac{dp_{c}}{d\xi},\label{eq_27}\\
&  -M\dfrac{dp_{c}}{d\xi}+u_{c}\dfrac{dp_{c}}{d\xi}+3p_{c}\dfrac{du_{c}}{d\xi}=0, \label{eq_28} \\
&  -M\dfrac{dn_{b}}{d\xi}+\frac{d(n_{b}u_{b})}{d\xi}=0,\label{eq_29}\\
&  -M\dfrac{du_{b}}{d\xi}+u_{b}\dfrac{du_{b}}{d\xi}=\dfrac{d\phi}{d\xi}%
-\frac{\theta_{b,h}}{n_{b}}\dfrac{dp_{b}}{d\xi},\label{eq_30}\\
&  -M\dfrac{dp_{b}}{d\xi}+u_{b}\dfrac{dp_{b}}{d\xi}+3p_{b}\dfrac{du_{b}}{d\xi
}=0, \label{eq_31} \\
\dfrac{d^{2}\phi}{d\xi^{2}}=& -\left(  1+\rho_{h,c}+\rho_{b,c}\right)  +n_{c}+\rho_{b,c}
n_{b} \nonumber\\
& +\rho_{h,c}\left(  1-\frac{\phi}{\kappa-\tfrac{3}{2}}\right)
^{-\kappa+1/2}, \label{eq_32}%
\end{align}%
It is assume that the equilibrium state is reached at
infinities ($\xi\rightarrow\pm\infty$). 
We then integrate the above fluid equations, 
apply the boundary conditions $n_{c}=1$, $p_{c}=1$, $u_{c}=0$, $n_{b}=1$,
$p_{b}=1$, $u_{b}=V_{b}$ and $\phi=0$ at infinities, and derive
\begin{align}
&  u_{c}=M[1-(1/n_{c})],\label{eq_33}\\
&  u_{c}={M-(M}^{2}{+2\phi-3n_{c}^{2}\theta_{c,h}+3\theta_{c,h}})^{1/2},\label{eq_34}\\
&  u_{b}=M[1-(1/n_{b})(1-V_{b}/M)],\label{eq_35}\\
&  u_{b}=M-[(M-V_{b})^{2}+2\phi-3n_{b}^{2}\theta_{b,h}+3\theta_{b,h}
]^{1/2},\label{eq_36}\\
&
\begin{array}
[c]{cc}%
p_c=n_c^{3},\text{ \ } & p_{b}=n_{b}^{3}.
\end{array}
\label{eq_37}%
\end{align}
Combining Eqs. (\ref{eq_33})--(\ref{eq_37}) and solving them lead to the
following solutions for the cool electron density, $n_{c}$, and the electron beam density, ${n}_{b}$, respectively:
\begin{align}
n_{c}   = & \frac{1}{2\sqrt{3\theta_{c,h}}}\left[  {2\phi+}({M+\sqrt
{3{\theta_{c,h}}}})^{2}\right]  ^{1/2}\nonumber\\
&  \pm\frac{1}{2\sqrt{3\theta_{c,h}}}{\left[  {2\phi+({M-\sqrt{3{\theta_{c,h}%
}}}})^{2}\right]  }^{1/2}, \label{eq_40} 
\end{align}
\begin{align}
{n}_{b}   = & \frac{1}{2\sqrt{3\theta_{b,h}}}\left[  {2\phi}+(M-V_{b}%
+\sqrt{3\theta_{b,h}})^{2}\right]  ^{1/2}\nonumber\\
&  \pm\frac{1}{2\sqrt{3\theta_{b,h}}}\left[  {2\phi}+(M-V_{b}-\sqrt
{3\theta_{b,h}})^{2}\right]  ^{1/2}. \label{eq_41}%
\end{align}
Eq. (\ref{eq_40}) is exactly the same as Eq.~(29) in Ref.~\onlinecite{Danehkar2011}. 
Taking the boundary conditions $n_{c}=n_{b}=1$ at $\phi=0$, the negative sign
must be used in equations (\ref{eq_40}) and (\ref{eq_41}). Furthermore, the
cool electrons and the electron beam are supersonic at $M>\sqrt
{3\theta_{c,h}}$ and $M>V_{b}+\sqrt{3\theta_{b,h}}$, while the hot electrons
are subsonic at $M<1$.
Considering the reality conditions of the density variables, the limits on the electrostatic
potential value become $\phi>\phi_{\lim(-)}=-\frac{1}{2}({M-}\sqrt{3{\theta_{c,h}}})^{2}$ for $V_{b}\leqslant0$ (counter-propagation case), and $-\frac{1}{2}({M-V_{b}-}\sqrt{3\theta_{b,h}})^{2}$ for $V_{b}>0$ (co-propagation case), which are associated with negatively polarized electrostatic solitary wave structures.

Substituting equations (\ref{eq_40})--(\ref{eq_41}) into the Poisson's
equation, multiplying the resulting equation by $d\phi/d\xi$, integrating and
taking into account the conditions at infinities ($d\phi/d\xi\rightarrow0$)
yield a pseudo-energy balance equation%
\begin{equation}
\frac{1}{2}\left(  \frac{d\phi}{d\xi}\right)  ^{2}+\Psi(\phi)=0, \label{eq_42}%
\end{equation}
where the Sagdeev pseudopotential $\Psi(\phi)$ is given by
\begin{align}
\Psi(\phi)=  &  \rho_{h,c}\left[  1-\left(  1+\frac{\phi}{-\kappa+\tfrac{3}%
{2}}\right)  ^{-\kappa+3/2}\right] \nonumber\\
&  +(1+\rho_{h,c}+\rho_{b,c})\phi\nonumber\\
&  +\frac{1}{6\sqrt{3{\theta_{c,h}}}}\left[  ({M+}\sqrt{3{\theta_{c,h}}}%
)^{3}-{{({M-}\sqrt{3{\theta_{c,h}}})^{3}}}\right. \nonumber\\
&  -({2\phi+}[{M+}\sqrt{3{\theta_{c,h}}}]^{2})^{3/2}\nonumber\\
&  \left.  +{({2\phi+[{M-}\sqrt{3{\theta_{c,h}}}]^{2}}})^{3/2}\right]
\nonumber\\
&  +\frac{\rho_{b,c}}{6\sqrt{3\theta_{b,h}}}\left[  (M-V_{b}{+}\sqrt
{3\theta_{b,h}})^{3}\right. \nonumber\\
&  -{{(M-V_{b}{-}\sqrt{3\theta_{b,h}})^{3}}}\nonumber\\
&  -({2\phi+}[M-V_{b}{+}\sqrt{3\theta_{b,h}}]^{2})^{3/2}\nonumber\\
&  \left.  +{({2\phi+[M-V_{b}{-}\sqrt{3\theta_{b,h}}]^{2}}})^{3/2}\right]  .
\label{eq_43}%
\end{align}
In the absence of the electron beam ($\rho_{b,c}\rightarrow0$), we
precisely recover Eq.~(33) in Ref.~\onlinecite{Danehkar2011}.

\section{Existence conditions for electron-acoustic solitons\label{existence}}

To maintain solitary waves, the origin at $\phi=0$ must be a root and a local
maximum of $\Psi$ in Eq. (\ref{eq_43}), so $\Psi(\phi)=0$, $\Psi^{\prime}%
(\phi)=0$ and $\Psi^{\prime\prime}(\phi)<0$ at $\phi=0$, where primes denote
derivatives with respect to $\phi$. As the first two constraints are
satisfied, the condition for the lower limit becomes $F_{1}(M)=-\Psi
^{\prime\prime}(\phi)|_{\phi=0}>0$. This yields the following equation for
the minimum value for the Mach number $M$:
\begin{align}
F_{1}(M)   = & \frac{\rho_{h,c}(\kappa-\frac{1}{2})}{\kappa-\tfrac{3}{2}}%
-\frac{1}{M^{2}-3\theta_{c,h}}\nonumber\\
&  -\frac{\rho_{b,c}}{(M-V_{b})^{2}-3\theta_{b,h}}. \label{eq_44}%
\end{align}
Eq. (\ref{eq_44}) constrains the lower limit for the Mach number, $M_{1}%
(\kappa,\rho_{h,c},\theta_{c,h},\rho_{b,c},\theta_{b,h},V_{b})$. 

An upper limit for $M$ is obtained when the density profile reaches the
complex value at $\phi_{\lim(-)}=-\frac{1}{2}\left(  {M-}\sqrt{3{\theta_{c,h}%
}}\right)  ^{2}$ for $V_{b}\leqslant0$ (counter-propagation case) and $\phi_{\lim(-)}=-\frac{1}{2}\left(
M-V_{b}{-}\sqrt{3{\theta_{b,h}}}\right)  ^{2}$ for $V_{b}>0$ (co-propagation case). Thus, the
largest soliton amplitude satisfies $F_{2}(M)=\Psi(\phi)|_{\phi=\phi_{\lim
(-)}}>0$. This yields the following equation for the upper limit in $M$ in the case of \textit{counter-propagating} ($V_{b}\leqslant0$),%
\begin{align}
F_{2}(M)  = & -\tfrac{1}{2}(1+\rho_{h,c}+\rho_{b,c})({M-}\sqrt{3{\theta_{c,h}}%
})^{2}\nonumber\\
&  -\tfrac{4}{3}M^{3/2}\left(  3\theta_{c,h}\right)  ^{1/4}\nonumber\\
&  +\rho_{h,c}\left[  1-\left(  1+\frac{[{M-}\sqrt{3{\theta_{c,h}}}]^{2}%
}{2\kappa-3}\right)  ^{-\kappa+3/2}\right]  \nonumber\\
&  +M^{2}+\theta_{c,h}+\rho_{b,c}(M-V_{b})^{2}{+}\rho_{b,c}\theta
_{b,h},\nonumber\\
&  +\frac{\rho_{b,c}}{6\sqrt{3\theta_{b,h}}}\left[  {({[M-V_{b}{-}%
\sqrt{3\theta_{b,h}}]^{2}}}\right.  \nonumber\\
&  {-{[{M-}\sqrt{3{\theta_{c,h}}}]^{2}}})^{3/2}-([M-V_{b}{+}\sqrt
{3\theta_{b,h}}]^{2}\nonumber\\
&  \left.  -{[{M-}\sqrt{3{\theta_{c,h}}}]^{2}})^{3/2}\right],  \label{eq_45}%
\end{align}
and the following equation for the upper limit in $M$ in the case of \textit{co-propagating} ($V_{b}>0$),%
\begin{align}
F_{2}(M)   = & -\tfrac{1}{2}(1+\rho_{h,c}+\rho_{b,c})(M-V_{b}{-}\sqrt
{3{\theta_{b,h}}})^{2}\nonumber\\
&  +M^{2}+\theta_{c,h}+\rho_{b,c}\theta_{b,h}+\rho_{b,c}\left(  M-V_{b}%
\right)  ^{2}\nonumber\\
&  -\tfrac{4}{3}\rho_{b,c}\left(  M-V_{b}\right)  ^{3/2}\left(  3\theta
_{b,h}\right)  ^{1/4}.\nonumber\\
&  +\rho_{h,c}\left[  1-\left(  1+\frac{[M-V_{b}{-}\sqrt{3{\theta_{b,h}}}%
]^{2}}{2\kappa-3}\right)  ^{-\kappa+3/2}\right]  \nonumber\\
&  +\frac{1}{6\sqrt{3{\theta_{c,h}}}}\left[  {({[{M-}\sqrt{3{\theta_{c,h}}%
}]^{2}}}\right.  \nonumber\\
&  {-[M-V_{b}{-}\sqrt{3{\theta_{b,h}}}]^{2}})^{3/2}-([{M+}\sqrt{3{\theta
_{c,h}}}]^{2}\nonumber\\
&  \left.  -[M-V_{b}{-}\sqrt{3{\theta_{b,h}}}]^{2})^{3/2}\right].
\label{eq_46}%
\end{align}
Solving equations (\ref{eq_45}) and (\ref{eq_46}) constrains the upper limit for
the Mach number, $M_{2}(\kappa,\rho_{h,c},\theta_{c,h},\rho_{b,c},\theta
_{b,h},V_{b})$. In the absence of the electron beam, Eqs. (\ref{eq_44})
and (\ref{eq_45}) recover precisely Eqs.~(34) and (36) in 
Ref.~\onlinecite{Danehkar2011}.%
\begin{figure}
[ptb]
\begin{center}
\includegraphics[width=3.1in]{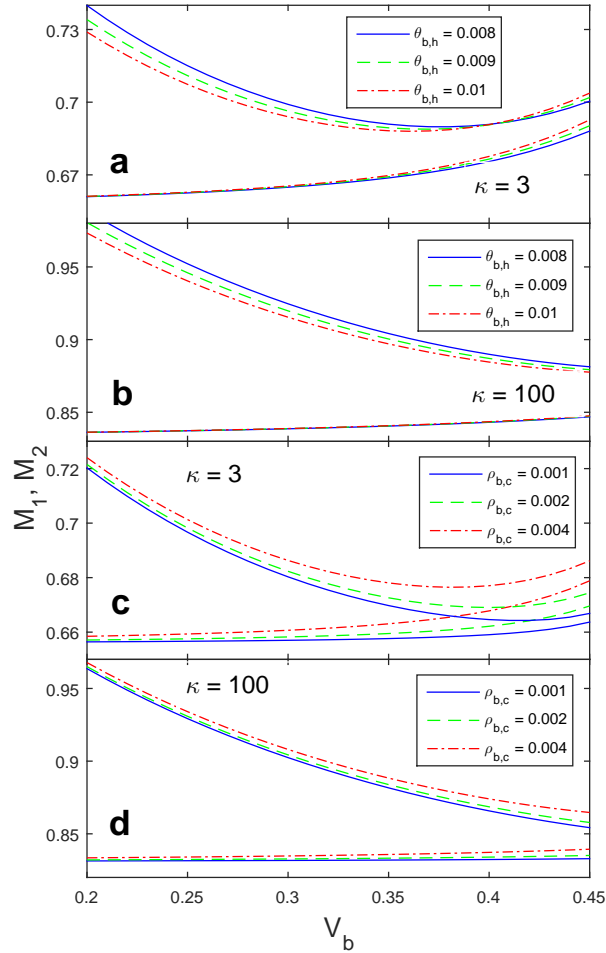}%
\caption{Variation of the lower limit for the Mach number $M_{1}$ (lower curves)
and the upper limit for the Mach number $M_{2}$ (upper curves) with the positive normalized
beam speed $V_{b}$ for different values of the beam-to-hot
electron temperature ratio $\theta_{b,h}$ (a-b) and different values of the
beam-to-cool electron number density ratio $\rho_{b,c}$ (c-d). Solitons
may exist for values of the Mach number $M$ in the region between the lower
and the upper curve(s) of the same style/color. (a-b) Curves: $\theta
_{b,h}=0.008$ (solid), $0.009$ (dashed), and $0.01$ (dot-dashed). Here, we
have taken: (a) $\kappa=3$ and (b) $\kappa=100$. (c-d) Curves: $\rho
_{b,c}=0.001$ (solid), $0.002$ (dashed), and $0.004$ (dot-dashed). Here, we
have taken: (c) $\kappa=3$ and (d) $\kappa=100$. The remaining parameters are
$\rho_{h,c}=1.5$, $\rho_{b,c}=0.008$ and $\theta_{c,h}=\theta_{b,h}=0.01$,
unless values are specified.}%
\label{fig2}%
\end{center}
\end{figure}
\begin{figure}[ptb]
\begin{center}
\includegraphics[width=3.1in]{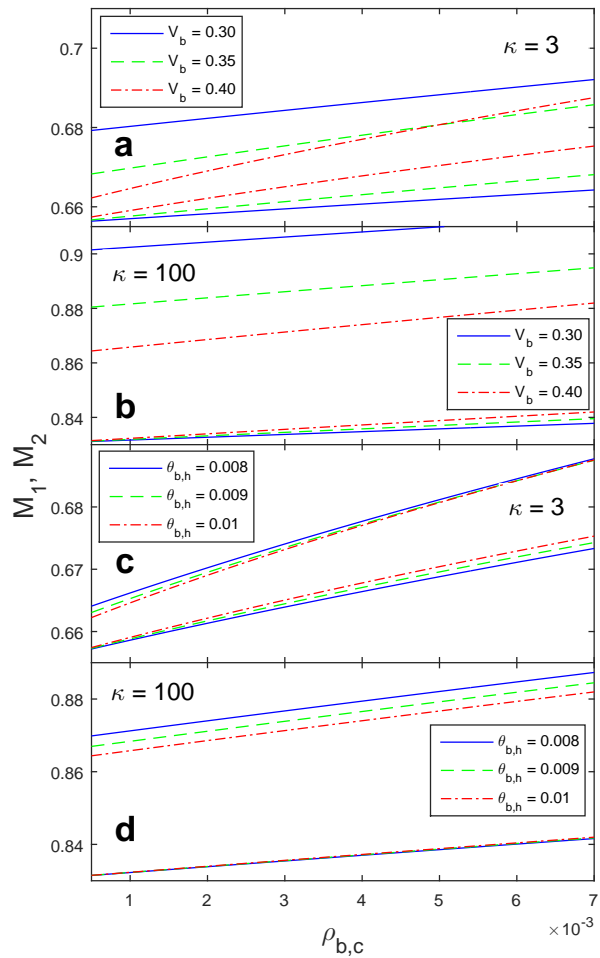}%
\caption{Variation of the lower limit for the Mach number $M_{1}$ (lower curves)
and the upper limit for the Mach number $M_{2}$ (upper curves) with the beam-to-cool electron
number density ratio $\rho_{b,c}$ for different values of the positive
normalized beam speed $V_{b}$ (a-b) and the beam-to-hot
electron temperature ratio $\theta_{b,h}$ (c-d). Solitons may exist for values
of the Mach number $M$ in the region between the lower and the upper curve(s)
of the same style/color. (a-b) Curves: $V_{b}=0.3$ (solid), $0.35$ (dashed),
and $0.4$ (dot-dashed). Here, we have taken: (a) $\kappa=3$ and (b)
$\kappa=100$. (c-d) Curves: $\theta_{b,h}=0.008$ (solid), $0.009$ (dashed),
and $0.01$ (dot-dashed). Here, we have taken: (c) $\kappa=3$ and (d)
$\kappa=100$. The remaining parameters are $\rho_{h,c}=1.5$, $V_{b}=0.4$ and
$\theta_{c,h}=\theta_{b,h}=0.01$, unless values are specified.}%
\label{fig3}%
\end{center}
\end{figure}
\begin{figure}[ptb]
\begin{center}
\includegraphics[width=3.1in]{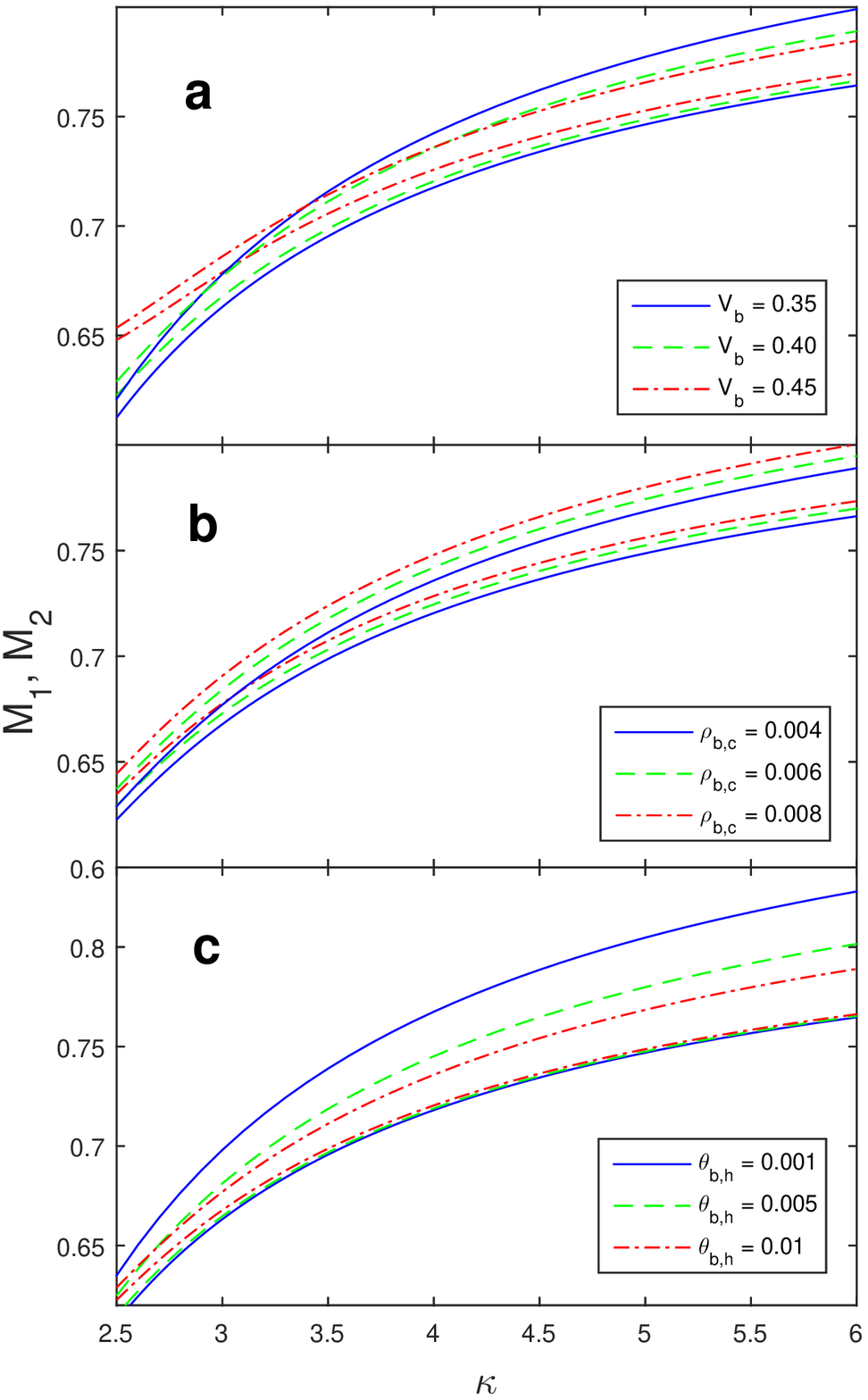}%
\caption{Variation of the lower limit for the Mach number $M_{1}$ (lower curves)
and the upper limit for the Mach number $M_{2}$ (upper curves) with the suprathermality parameter
$\kappa$ for different values of the positive normalized beam 
speed $V_{b}$ (upper panel), the beam-to-cool electron number density
ratio $\rho_{b,c}$ (middle panel) and the beam-to-hot electron temperature
ratio $\theta_{b,h}$ (bottom panel). Solitons may exist for values of the Mach
number $M$ in the region between the lower and upper curves of the same
style/color. Upper panel: $V_{b}=0.35$ (solid curve), $0.40$ (dashed), and
$0.45$ (dot-dashed). Here, we have taken $\rho_{h,c}=1.5$, $\theta
_{c,h}=\theta_{b,h}=0.01$ and $\rho_{b,c}=0.004$. Middle panel: $\rho
_{b,c}=0.004$ (solid curve), $0.006$ (dashed), and $0.008$ (dot-dashed). Here,
we have taken $\rho_{h,c}=1.5$, $\theta_{c,h}=\theta_{b,h}=0.01$ and
$V_{b}=0.4$. Lower panel: $\theta_{b,h}=0.001$ (solid curve), $0.005$
(dashed), and $0.01$ (dot-dashed). Here, we have taken $\rho_{h,c}=1.5$,
$\rho_{b,c}=0.004$, $\theta_{c,h}=0.01$ and $V_{b}=0.4$.}%
\label{fig4}%
\end{center}
\end{figure}

The soliton existence regions are shown in Figs. \ref{fig2}--\ref{fig4} for
different parameters of the electron beam and different values of the suprathermality (described by $\kappa$) of the hot electrons. Solitary structures of the electrostatic potential may
occur in the range $M_{1}<M<M_{2}$, which depends on the parameters $\kappa$,
$\rho_{h,c}$, $\theta_{c,h}$, $\rho_{b,c}$, $\theta_{b,h}$ and $V_{b}$.
Furthermore, we assumed that the cool electrons are supersonic (${M>}%
\sqrt{3{\theta_{c,h}}}$), as well as the electron beam (${M>V_{b}%
+\sqrt{3{\theta_{b,h}}}}$),
whereas the hot electrons are subsonic (${M<1}$).

Figure \ref{fig2} shows the existence domain of electron-acoustic solitary
waves in two opposite cases: in (a) a very high suprathermality ($\kappa=3$), and (b) a very low suprathermality ($\kappa=100$). The normalized beam speed is positive ($V_{b}>0$),
which means that the soliton \textit{co-propagates} with the
electron beam. We see that the existence domain of the Mach number
becomes narrower with strong suprathermality and higher values of the normalized beam speed. Comparing the two frames (a) and (b), or (c) and (d) in Fig.
\ref{fig2}, we notice that low values of $\kappa$ impose that the soliton
propagates at lower Mach number range. From Fig. \ref{fig2}a-b, it can be seen
that increasing the electron beam thermal pressure ($\theta_{b,h}$)
shrinks the soliton existence region. Finally, we note that lower values of
the beam-to-cool electron number density ratio ($\rho_{b,c}$; also see
Fig. \ref{fig3} for this effect) shrink the permitted soliton region, nearly down to
nil, for very high $V_{b}$ ($\geqslant0.5$) and strong suprathermality (low
$\kappa$).

As seen in Figs.~\ref{fig2} and \ref{fig3}, the existence region becomes
narrower for lower values of $\rho_{b,c}$ and $\kappa$. This is especially
visible in Fig. \ref{fig2}a,c; an increase in $\rho_{b,c}$ broadens the
permitted region of the Mach number. It is in contrast to increasing the
hot-to-cool electron number density ratio ($\rho_{h,c}$), which shrinks down
the existence region \citep{Danehkar2011}. We know that a reduction in the cool
electron density (increasing $\rho_{h,c}$) was responsible for shrinking the
existence region. Moreover, we also notice that a reduction in the electron beam density (decreasing $\rho_{b,c}$) has a similar effect in the soliton
existence region. Similarly, Fig. \ref{fig3} (also see Fig. \ref{fig2})
shows that higher values of the normalized beam speed $V_{b}$ and
the temperature ratio $\theta_{b,h}$ shrink the permitted region in a strong
suprathermality (low $\kappa$).

Figure \ref{fig4} demonstrates the effect of the hot electron $\kappa$-distribution. 
The acoustic limits ($M_{1}$ and $M_{2}$) decrease rapidly as
approaching the limiting value $\kappa\rightarrow3/2$. Moving toward a
Maxwellian distribution ($\kappa\rightarrow\infty$) broadens the permitted
range of the Mach number. The result is similar to the trend in Figs.
\ref{fig2} and \ref{fig3}; stronger suprathermality imposes that solitons
occur in compressed ranges of the Mach number. It is quite similar to what we
found in our previous model without the electron beam \citep{Danehkar2011}.

\begin{figure}
[ptb]
\begin{center}
\includegraphics[width=3.1in]{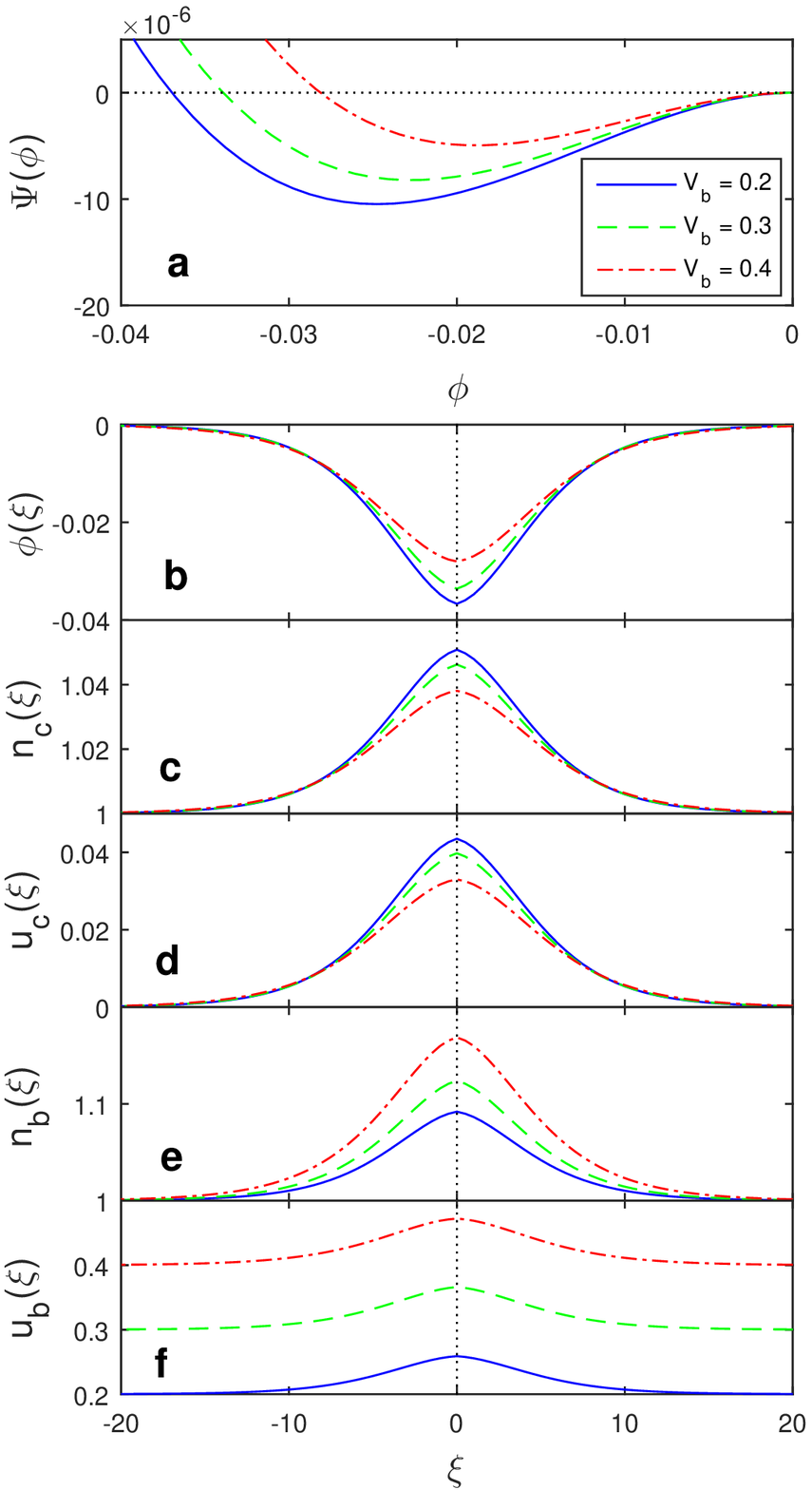}
\caption{(a) The pseudopotential $\Psi(\phi)$ of electron-acoustic solitons and the
associated solutions: (b) electric potential pulse $\phi$, (c) density $n$ and
(d) velocity $u$ of the cool electron fluid, (e) density $n_{b}$ and (f)
velocity $u_{b}$ of the electron beam are depicted versus position $\xi$, 
for different values of the positive normalized beam speed $V_{b}$. We have
taken: $V_{b}=0.2$ (solid curve), $0.3$ (dashed curve), and $0.4$ (dot-dashed
curve). The other parameter values are: $\rho_{h,c}=1$, $\rho_{b,c}=0.008$,
$\theta_{c,h}=\theta_{b,h}=0.01$, $\kappa=4.0$ and $M=0.9$.}%
\label{fig5}%
\end{center}
\end{figure}
\begin{figure}
[ptb]
\begin{center}
\includegraphics[width=3.2in]{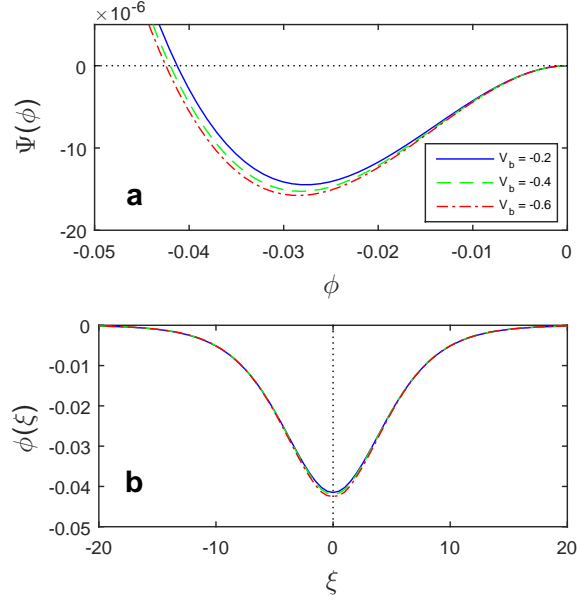}%
\caption{(a) The pseudopotential $\Psi(\phi)$ of electron-acoustic solitons and the associated solutions: (b) electric potential pulse $\phi$ is depicted versus position $\xi$, for different values of the negative normalized beam speed $V_{b}$. We have taken: $V_{b}=-0.2$ (solid curve), $-0.4$
(dashed curve), and $-0.6$ (dot-dashed curve). The other parameter values are:
$\rho_{h,c}=1$, $\rho_{b,c}=0.008$, $\theta_{c,h}=\theta_{b,h}=0.01$,
$\kappa=4.0$ and $M=0.9$.}%
\label{fig8}%
\end{center}
\end{figure}

\begin{figure}[ptb]
\begin{center}
\includegraphics[width=3.2in]{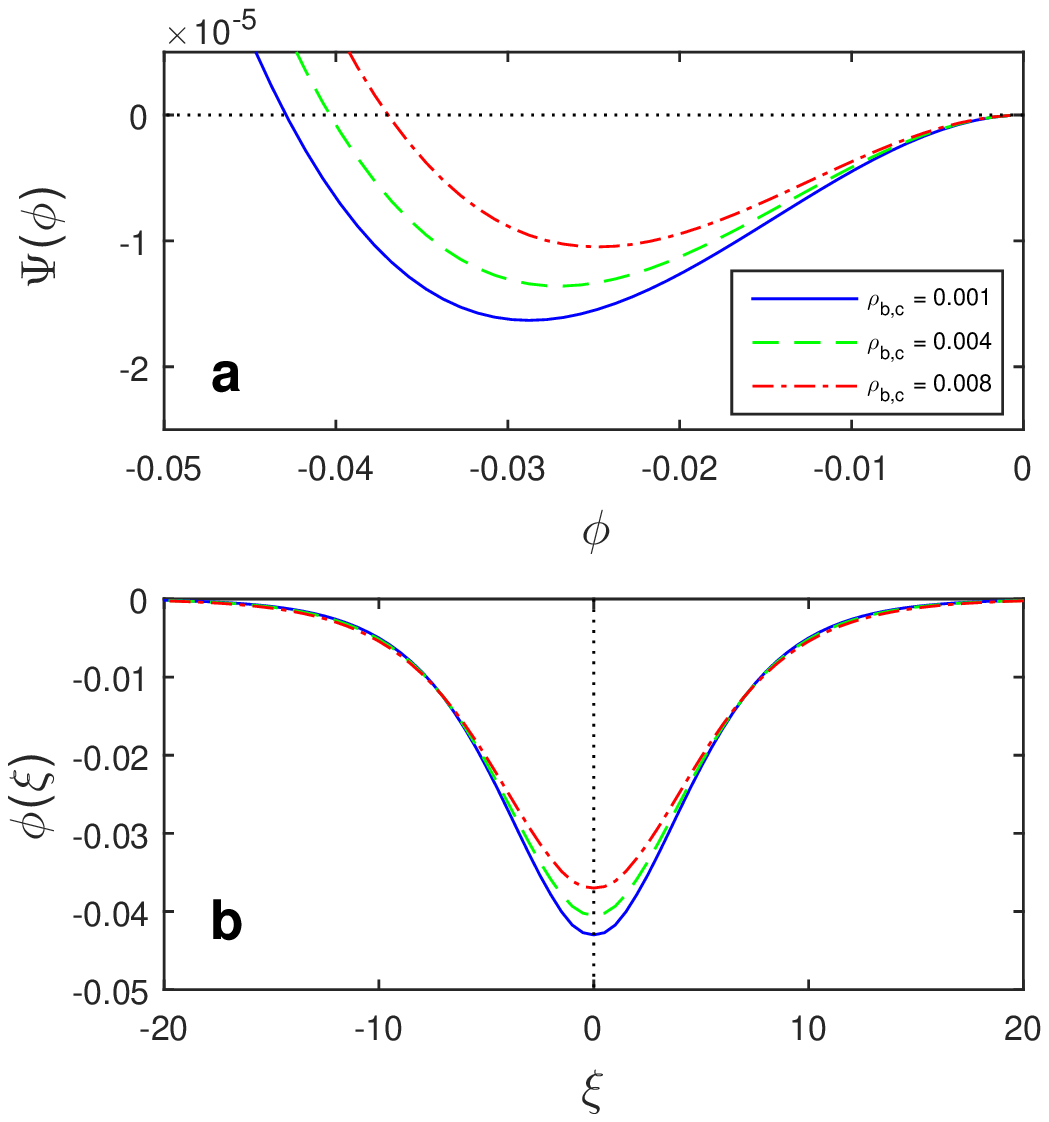}%
\caption{(a) The pseudopotential $\Psi(\phi)$ of electron-acoustic solitons and the associated solutions: (b) electric potential pulse $\phi$ is depicted versus position $\xi$, for different values of the beam-to-cool electron
number density ratio $\rho_{b,c}$. From bottom to top: $\rho_{b,c}=0.001$
(solid curve); $0.004$ (dashed curve); $0.008$ (dot-dashed curve). Here
$\rho_{h,c}=1$, $\theta_{c,h}=\theta_{b,h}=0.01$, $V_{b}=0.2$, $\kappa=4.0$
and $M=0.9$.}%
\label{fig6}%
\end{center}
\end{figure}

We also notice that varying the negative normalized beam speed
($V_{b}<0$) trivially change the lower limit $M_{1}$ and the upper limit
$M_{2}$ under the excitation conditions $\rho_{b,c} V_{b} \ll 1 $ and $\theta_{b,h}\ll0.1$ (figures not shown here). Similarly, the electron beam thermal
pressure ($\theta_{b,h}$) and the beam-to-cool electron number density
ratio ($\rho_{b,c}$) have trivial effects on the existence domain (not shown
here) when the normalized beam speed is negative, in contrast to
what we saw in Fig.~\ref{fig2}a,b. Therefore, the existence domain of
electron-acoustic solitary waves does not affect largely when the soliton
\textit{counter-propagates} with the electron beam. %

\section{Nonlinear electron-acoustic wave structures \label{investigation}}

\begin{figure}[ptb]
\begin{center}
\includegraphics[width=3.2in]{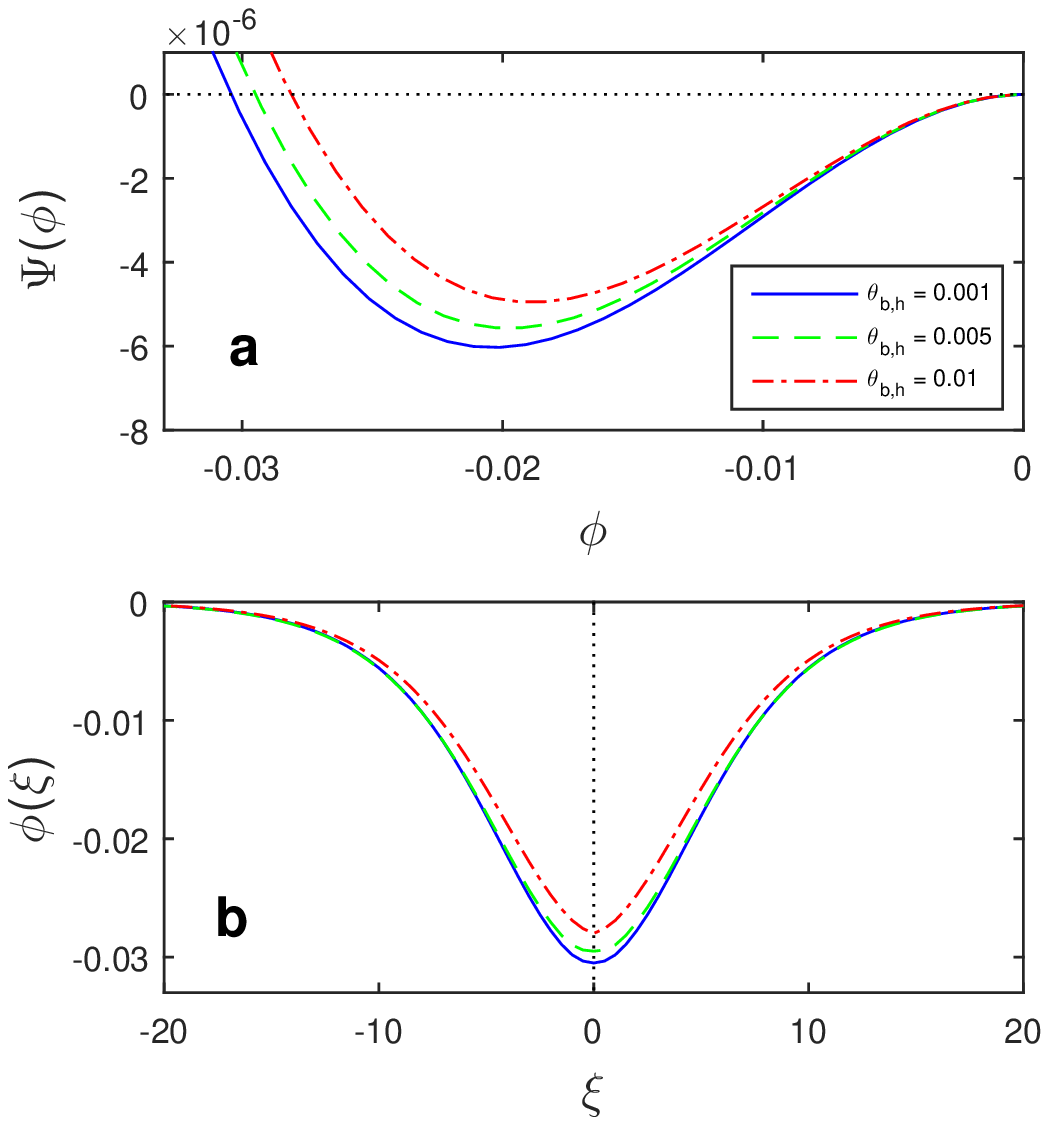}%
\caption{(a) The pseudopotential $\Psi(\phi)$ of electron-acoustic solitons and the associated solutions: (b) electric potential pulse $\phi$ is depicted versus position $\xi$, for different values of the beam-to-hot electron
temperature ratio $\theta_{b,h}$. From bottom to top: $\theta_{b,h}=0.001$
(solid curve); $0.005$ (dashed curve); $0.01$ (dot-dashed curve). Here
$\rho_{h,c}=1$, $\rho_{b,c}=0.008$, $\theta_{c,h}=0.01$, $V_{b}=0.4$,
$\kappa=4.0$ and $M=0.9$.}%
\label{fig7}%
\end{center}
\end{figure}

To study the localized structures of large-amplitude EAWs,
we have solved Eq.~(\ref{eq_42}) via numerical integration for various plasma
parameters, which allow us to observe their effects on the nonlinear wave
structures. We have found only negatively polarized solitons, in contrast to
what found in Ref.~\onlinecite{Berthomier2000}, as the $\kappa$-distribution may not facilitate reverse polarity of electrostatic solitary waves. 
To have positive polarity in our fluid model, a mobile positive charge such as ion and positron is required to support the inertia for acoustic oscillations, while the suprathermal hot electrons provide the wave restoring force. However, the ion inertia propagates electrostatic waves in a slow-acoustic mode \citep{Danehkar2009}. Thus, mobile positrons (or electron holes) may provide the inertia to have positive polarity in a fast-acoustic mode similar to the negatively polarized electron-acoustic solitons \citep{Danehkar2017}. 

Figure \ref{fig5} shows the variation of the pseudopotential $\Psi(\phi)$ with
the normalized potential $\phi$, for different values of the positive normalized beam speed $V_{b}$ (keeping $\rho_{h,c}=1$, $\rho_{b,c}=0.008$,
$\theta_{c,h}=\theta_{b,h}=0.01$, $\kappa=4.0$ and Mach number $M=0.9$, all
fixed). The pulse amplitude $\left\vert \phi_{m}\right\vert $ decreases with
increasing $V_{b}$, which are in agreement with Refs.~\onlinecite{Devanandhan2011b,Singh2011} and \onlinecite{Singh2016}.
We algebraically determined the fluid density (Fig.
\ref{fig5}c) and velocity disturbances (Fig. \ref{fig5}d) of the cool
electrons and the electron beam (Fig. \ref{fig5}e,f). It is noticeable that
an increase in positive $V_{b}$ (co-propagation) compresses the disturbances of $u_{c}$, $n_{c}$, $n_{b}$ and
$u_{b}-V_{b}$. 
Similarly, Figure \ref{fig8} depicts the variation of the
pseudopotential $\Psi(\phi)$ for different values of the negative normalized beam speed $V_{b}$. It is seen that the pulse amplitude $\left\vert
\phi_{m}\right\vert $ 
is slightly altered with a change in negative $V_{b}$ (counter-propagation), in contrast to what we saw in Fig.~\ref{fig5}.

\begin{figure}[ptb]
\begin{center}
\includegraphics[width=3.2in]{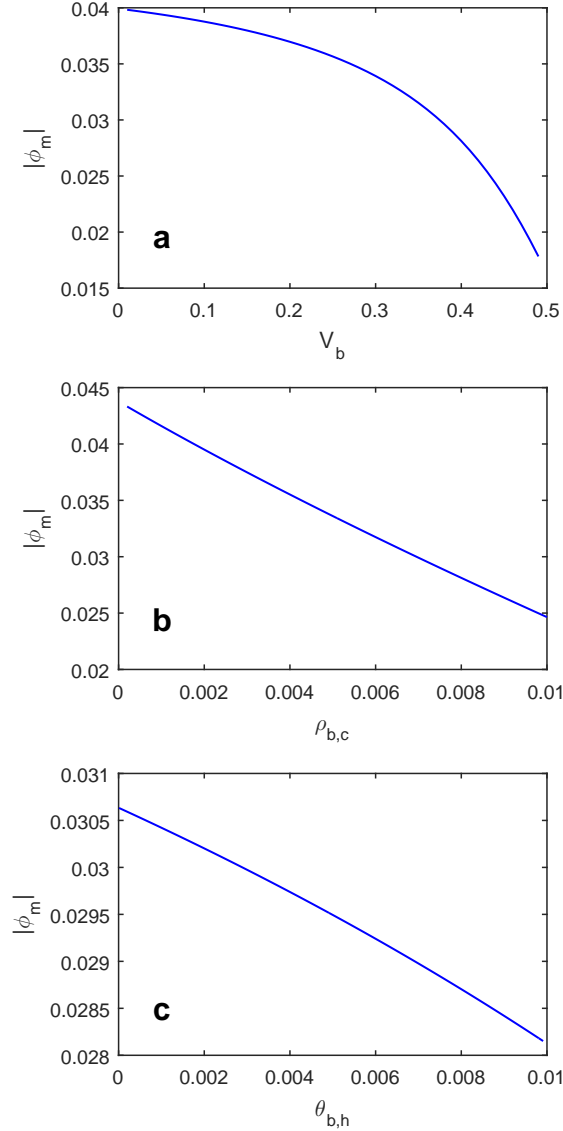}%
\caption{Variation of the absolute pulse amplitude $|\phi_{m}|$ of electron-acoustic solitons versus (a) the positive normalized beam speed $V_{b}$, 
(b) the beam-to-cool electron number density ratio $\rho_{b,c}$, and (c) the beam-to-hot electron temperature ratio $\theta_{b,h}$.  
The remaining parameters are
$\rho_{h,c}=1$, $\rho_{b,c}=0.008$, $\theta_{c,h}=0.01$, $\theta_{b,h}=0.01$, $V_{b}=0.4$,
$\kappa=4.0$ and $M=0.9$, unless values are specified.}%
\label{fig9}%
\end{center}
\end{figure}

\begin{figure}[ptb]
\begin{center}
\includegraphics[width=3.2in]{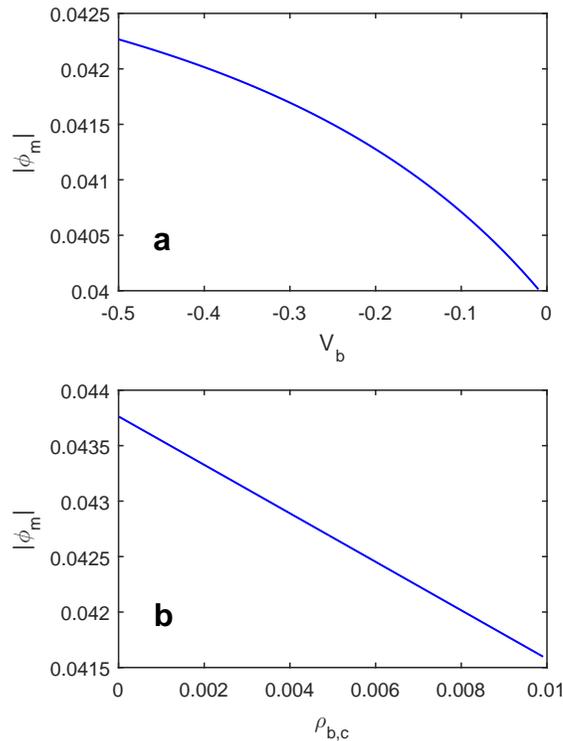}%
\caption{Variation of the absolute pulse amplitude $|\phi_{m}|$ of electron-acoustic solitons versus (a) the negative normalized beam speed $V_{b}$ and 
(b) the beam-to-cool electron number density ratio $\rho_{b,c}$. 
The remaining parameters are
$\rho_{h,c}=1$, $\rho_{b,c}=0.008$, $\theta_{c,h}=0.01$, $\theta_{b,h}=0.01$, $V_{b}=-0.4$,
$\kappa=4.0$ and $M=0.9$, unless values are specified.}%
\label{fig10}%
\end{center}
\end{figure}

Figure \ref{fig6} shows the variation of the pseudopotential $\Psi(\phi)$ for
different values of the beam-to-cool electron number density ratio
$\rho_{b,c}$. Both the root and the depth of the Sagdeev potential well
increase with decreasing $\rho_{b,c}=n_{b,0}/n_{c,0}$. This means that
increasing the cool electron density and/or decreasing the electron beam density increases the negative polarity solitary waves.

Figure \ref{fig7} shows the thermal effect of the electron beam through $\theta_{b,h}=T_{b}%
/T_{h}$, which agrees with Refs.~\onlinecite{Devanandhan2011b} and ~\onlinecite{Singh2016}.
We see that the soliton excitations are
amplified with either an increase in the suprathermal hot electron temperature
($T_{h}$) or a decrease in the electron beam temperature ($T_{b}$). Therefore,
thermal effects of the hot electrons relative to the electron beam amplify
the electric potential at a fixed Mach number (similar to what found for the
hot relative to the cool electrons via $\theta_{c,h}=T_{c}/T_{h}$ in Ref.~\onlinecite{Danehkar2011}).

Figure \ref{fig9} summarizes the behavior of the negative polarity EAWs co-propagating with the beam speed ($V_{b}>0$). 
It can be seen that the maximum absolute pulse amplitude $|\phi_{m}|$ of negative electric potentials is significantly reduced by increasing the normalized beam speed $V_{b}$ and
the beam-to-cool electron number density ratio $\rho_{b,c}$ (see Fig.~\ref{fig9}a,b). 
However, an increase in the beam-to-hot electron temperature ratio $\theta_{b,h}$ slightly decreases the maximum absolute pulse amplitude $|\phi_{m}|$ 
of negative electric potentials (see Fig.~\ref{fig9}c). Similarly, in Figure \ref{fig10}, we have depicted the variation of the absolute pulse amplitude of negative polarity EAWs counter-propagating 
with the electron beam ($V_{b}<0$) versus the normalized beam speed $V_{b}$ and 
the beam-to-cool electron number density ratio $\rho_{b,c}$.  It is seen that the absolute pulse amplitude $|\phi_{m}|$ of negatively polarized EAWs slightly grows with an increase 
in the absolute normalized speed $|V_{b}|$ of the counter-propagating electron beam. 
Moreover, Figure~\ref{fig10}(b) shows that the absolute pulse amplitude of negative polarity solitons is slightly reduced by increasing the beam-to-cool electron number density ratio $\rho_{b,c}$ in the counter-propagation case, 
which is dissimilar to the large amplification in the co-propagation case shown in Fig.~\ref{fig9}(b). 
Under the trivially current ($\rho_{b,c} V_{b} \ll 1 $) and weakly damped ($\theta_{b,h}\ll0.1$) excitation conditions, varying the beam-to-hot electron temperature ratio $\theta_{b,h}$ insignificantly alters the maximum absolute pulse amplitude of negative electric potentials in the counter-propagation situation (figure not presented here).

\section{Conclusions \label{conclusion}}

We have studied the propagation of electron-acoustic
solitary waves in a collisionless, unmagnetized plasma model consisting of cool inertial background
electrons, cool inertial beam electrons, $\kappa$-distributed suprathermal
hot electrons and stationary ions.

We have obtained a linear dispersion relation for the small-amplitude EAWs and determined
the effects of electron beam properties on the dispersion relation,
through the beam-to-cool electron population ratio $\rho_{b,c}$ and the
normalized beam speed $V_{b}$. It is found that the phase speed
increases weakly with an increase in $\rho_{b,c}$ and $V_{b}$, while it is
largely decreased by increasing the suprathermality (decreasing $\kappa$ and
increasing $\rho_{h,c}$; see Ref.~\onlinecite{Danehkar2011}).

We have employed the Sagdeev's pseudopotential technique to study the nonlinear
dynamics of large-amplitude EAWs, and 
to determine the permitted parametric regions where allow solitons to propagate in
the plasma. The results of this study indicate that the existence domain of
solitons, which \textit{co-propagate} with the electron beam 
($V_{b}>0$), becomes narrower with an increase in the suprathermality (decreasing
the spectral index $\kappa$), increasing the normalized beam speed $V_{b}$,
decreasing the beam-to-cool electron population ratio $\rho_{b,c}$,
and increasing the beam-to-hot electron temperature ratio $\theta_{b,h}$. 
However, the \textit{counter-propagation} ($V_{b}<0$) slightly affects the existence domain of the nonlinear EAWs.

We have numerically solved the pseudo-energy balance equation (\ref{eq_42}) to
study effects of various plasma parameters on the nonlinear EAWs. It is
found that increasing the beam-to-cool electron density ratio $\rho_{b,c}$
largely decreases the electric potential amplitude in the co-propagation case ($V_{b}>0$; see Figs.~\ref{fig6} and \ref{fig9}), but slightly in the counter-propagation case ($V_{b}<0$; see Fig.~\ref{fig10}). Increasing 
the positive beam speed ($V_{b}>0$) significantly decreases 
the soliton electric potential amplitude (Figs.~\ref{fig5} and \ref{fig9}). However, a change 
in the negative beam speed ($V_{b}<0$) slightly alters the electric potential amplitude (Figs.~\ref{fig8} and \ref{fig10}) under the excitation conditions ($\rho_{b,c} V_{b} \ll 1 $ and $\theta_{b,h}\ll0.1$). 
We also see that decreasing the beam-to-hot electron
temperature ratio $\theta_{b,h}$ slightly amplifies the electric potential amplitude in the co-propagation case ($V_{b}>0$; see Figs.~\ref{fig7} and \ref{fig9}), but insignificantly alters the soliton pulse in the counter-propagation situation ($V_{b}<0$) under the excitation conditions.   
Some of these results are aligned with the previous findings by other authors \citep{Devanandhan2011b,Singh2011,Singh2016}. 

Our two-fluid model predicts only the propagation of negatively polarized EAWs. However, we
did not find any positive polarity electrostatic solitons. This is in
agreement with Ref.~\onlinecite{Devanandhan2011b}, but disagrees with 
Refs.~\onlinecite{Berthomier2000} and \onlinecite{Singh2011}. Apparently, the inclusion of $\kappa
$-distribution (or inertialess) electrons does not permit positive polarity
electrostatic solitons. 
Refs.~\onlinecite{Verheest2005} and \onlinecite{Cattaert2005} concluded
that positive polarity electrostatic solitary waves can be produced in a parameter range if
hot electron inertia is retained. In order to find solitons with a positive
polarity signature, it is imperative to retain hot electron inertia in
combination with the thermal effects \citep{Verheest2007}. However, it still
requires a high density of cool electrons to produce positive electrostatic
potential solitons \citep{Verheest2007}, larger than what are currently measured in
 observations. 
It is known that the ion inertia, combined with suprathermal electrons, can generate positive polarity electrostatic waves, but in the ion-acoustic mode, and propagate much slower than electrostatic solitary waves in the electron-acoustic mode (see e.g. \citep{Danehkar2009}). Alternatively, mobile cool positrons (or electron holes), together with suprathermal hot electrons,  may support positive polarity electrostatic waves with the propagation speed comparable to the negative polarity electron-acoustic solitons \citep{Danehkar2017}. 

The present study was motivated by electrostatic solitary waves observed in the BEN \citep{Dubouloz1991a}, the PCBL region \citep{Tsurutani1998}, the magnetopause \citep{Ergun2016}, and the magnetosheath \citep{Holmes2018} of the Earth's magnetosphere, where hot suprathermal electrons and magnetic field-aligned electron beams are found to be present. For example, considering the parameters of the BEN plasma ($T_{b} \sim 1$\,eV \citep{Schriver1989}, $T_{h} \sim 990$\,eV, and $n_c \sim n_h \sim 10$ cm$^{-3}$ \citep{Dubouloz1991a}), the peak-to-peak amplitude of the normalized electric field is calculated to be $E=-\partial \phi / \partial x \approx 7\times 10^{-3}$ (figures not shown) based on the electric potential pulse $\phi$ at $\theta_{b,h}=T_{b}/T_{h}=0.001$ with the physical parameters listed in Figure~\ref{fig7}. 
This corresponds to a peak-to-peak electric field of $E \approx 90$\,mV/m (potential $\phi$ and space $x$ are, respectively, unnormalized by $k_{B}T_{h}/e=9.9 \times 10^{5}$\,mV and $\lambda_{0}=\sqrt{\varepsilon_{0}k_{B}T_{h}/n_{c,0}e^{2}}=74$\,m), which is in agreement with the peak-to-peak electric filed of $\sim 100$\,mV/m observed in the BEN \citep{Dubouloz1991a}.
Similarly, the parameters of the magnetosphere plasma observed by the MMS mission \citep{Ergun2016,Holmes2018} ($T_{b} \sim 1$\,eV, $T_{h} \sim 1$\,keV, $n_b \sim 0.2$ cm$^{-3}$,  and $n_h \sim 30$ cm$^{-3}$) yield a peak-to-peak electric field of $E \approx 160$\,mV/m ($k_{B}T_{h}/e= 10^{6}$\,mV and $\lambda_{0}=43$\,m), which is roughly close to the large-amplitude, parallel electric fields of $\sim 100$\,mV/m measured at the magnetic reconnection region of the Earth's magnetopause \citep{Ergun2016}.

In conclusion, the parametric investigations presented in Sections \ref{existence} and \ref{investigation} suggest that the propagation of
electron-acoustic solitary waves, as well as the nonlinear dynamics of large-amplitude electrostatic wave structures, can be
altered by the electron beam, in the presence of suprathermal electrons. 
The results could have important  implications for the characteristics of nonlinear plasma waves observed in several space and astrophysical plasma environments such as the Earth's magnetosphere.

\ack 

The author was in receipt of a postgraduate studentship from the Department for Employment and
Learning (DEL) in Northern Ireland during the postgraduate study at the Queen's University Belfast, and held a Research Excellence Scholarship while at Macquarie University. It is a pleasure to acknowledge Dr. I. Kourakis and Prof. M. A. Hellberg for fruitful discussions.

\providecommand{\newblock}{}


\begin{thebibliography}{10}
\expandafter\ifx\csname url\endcsname\relax
  \def\url#1{{\tt #1}}\fi
\expandafter\ifx\csname urlprefix\endcsname\relax\def\urlprefix{URL }\fi
\providecommand{\eprint}[2][]{\url{#2}}

\bibitem{Watanabe1977}
{Watanabe} K and {Taniuti} T 1977 {\em \jpsj\/} {\bf 43} 1819

\bibitem{Ikezawa1981}
{Ikezawa} S and {Nakamura} Y 1981 {\em \jpsj\/} {\bf 50} 962--967

\bibitem{Mace1990}
{Mace} R~L and {Hellberg} M~A 1990 {\em \jplph\/} {\bf 43} 239--255

\bibitem{Hellberg2000}
{Hellberg} M~A, {Mace} R~L, {Armstrong} R~J, and {Karlstad} G 2000 {\em
  \jplph\/} {\bf 64} 433--443

\bibitem{Dubouloz1991}
{Dubouloz} N, {Pottelette} R, {Malingre} M, and {Treumann} R~A 1991 {\em
  \grl\/} {\bf 18} 155--158

\bibitem{Feldman1983}
{Feldman} W~C, {Anderson} R~C, {Bame} S~J, {Gary} S~P, {Gosling} J~T, {McComas}
  D~J, {Thomsen} M~F, {Paschmann} G, and {Hoppe} M~M 1983 {\em \jgr\/} {\bf 88}
  96--110

\bibitem{Bale1998}
{Bale} S~D, {Kellogg} P~J, {Larsen} D~E, {Lin} R~P, {Goetz} K, and {Lepping}
  R~P 1998 {\em \grl\/} {\bf 25} 2929--2932

\bibitem{Pottelette1999}
{Pottelette} R, {Ergun} R~E, {Treumann} R~A, {Berthomier} M, {Carlson} C~W,
  {McFadden} J~P, and {Roth} I 1999 {\em \grl\/} {\bf 26} 2629--2632

\bibitem{Cattell2002}
{Cattell} C, {Crumley} J, {Dombeck} J, {Wygant} J~R, and {Mozer} F~S 2002 {\em
  \grl\/} {\bf 29} 1065

\bibitem{Baluku2011}
{Baluku} T~K, {Hellberg} M~A, and {Mace} R~L 2011 {\em \jgra\/} {\bf 116}
  A04227

\bibitem{Gary1985}
{Gary} S~P and {Tokar} R~L 1985 {\em \phfl\/} {\bf 28} 2439--2441

\bibitem{Mace1999}
{Mace} R~L, {Amery} G, and {Hellberg} M~A 1999 {\em \phpl\/} {\bf 6} 44--49

\bibitem{Tokar1984}
{Tokar} R~L and {Gary} S~P 1984 {\em \grl\/} {\bf 11} 1180--1183

\bibitem{Lin1984}
{Lin} C~S, {Burch} J~L, {Shawhan} S~D, and {Gurnett} D~A 1984 {\em \jgr\/} {\bf
  89} 925--935

\bibitem{Ogilvie1984}
{Ogilvie} K~W, {Fitzenreiter} R~J, and {Scudder} J~D 1984 {\em \jgr\/} {\bf 89}
  10723--10732

\bibitem{Pottelette1990}
{Pottelette} R, {Malingre} M, {Dubouloz} N, {Aparicio} B, and {Lundin} R 1990
  {\em \jgr\/} {\bf 95} 5957--5971

\bibitem{Dubouloz1991a}
{Dubouloz} N, {Pottelette} R, {Malingre} M, {Holmgren} G, and {Lindqvist} P~A
  1991 {\em \jgr\/} {\bf 96} 3565--3579

\bibitem{Matsumoto1994}
{Matsumoto} H, {Kojima} H, {Miyatake} T, {Omura} Y, {Okada} M, {Nagano} I, and
  {Tsutsui} M 1994 {\em \grl\/} {\bf 21} 2915--2918

\bibitem{Cattell1998}
{Cattell} C, {Wygant} J, {Dombeck} J, {Mozer} F~S, {Temerin} M, and {Russell}
  C~T 1998 {\em \grl\/} {\bf 25} 857--860

\bibitem{Tsurutani1998}
{Tsurutani} B~T, {Arballo} J~K, {Lakhina} G~S, {Ho} C~M, {Buti} B, {Pickett}
  J~S, and {Gurnett} D~A 1998 {\em \grl\/} {\bf 25} 4117--4120

\bibitem{Ergun2016}
{Ergun} R~E, {Holmes} J~C, {Goodrich} K~A, {Wilder} F~D, {Stawarz} J~E,
  {Eriksson} S, {Newman} D~L, {Schwartz} S~J, {Goldman} M~V, {Sturner} A~P,
  {\em et~al.\/} 2016 {\em \grl\/} {\bf 43} 5626--5634

\bibitem{Holmes2018}
{Holmes} J~C, {Ergun} R~E, {Newman} D~L, {Wilder} F~D, {Sturner} A~P,
  {Goodrich} K~A, {Torbert} R~B, {Giles} B~L, {Strangeway} R~J, and {Burch} J~L
  2018 {\em \jgra\/} {\bf 123} 132--145

\bibitem{Berthomier2000}
{Berthomier} M, {Pottelette} R, {Malingre} M, and {Khotyaintsev} Y 2000 {\em
  \phpl\/} {\bf 7} 2987--2994

\bibitem{Mace2001}
{Mace} R~L and {Hellberg} M~A 2001 {\em \phpl\/} {\bf 8} 2649--2656

\bibitem{Sahu2004}
{Sahu} B and {Roychoudhury} R 2004 {\em \phpl\/} {\bf 11} 1947--1954

\bibitem{ElTaibany2005a}
{El-Taibany} W~F 2005 {\em \jgra\/} {\bf 110} A01213

\bibitem{ElTaibany2005b}
{El-Taibany} W~F and {Moslem} W~M 2005 {\em \phpl\/} {\bf 12} 032307

\bibitem{ElLabany2005}
{El-Labany} S~K, {El-Taibany} W~F, and {El-Abbasy} O~M 2005 {\em \phpl\/} {\bf
  12} 092304

\bibitem{Elwakil2007}
{Elwakil} S~A, {Zahran} M~A, and {El-Shewy} E~K 2007 {\em \physscr\/} {\bf 75}
  803--808

\bibitem{Devanandhan2011b}
{Devanandhan} S, {Singh} S~V, {Lakhina} G~S, and {Bharuthram} R 2011 {\em
  \npgeo\/} {\bf 18} 627--634

\bibitem{Singh2011}
{Singh} S~V, {Lakhina} G~S, {Bharuthram} R, and {Pillay} S~R 2011 {\em \phpl\/}
  {\bf 18} 122306

\bibitem{Lakhina2008a}
{Lakhina} G~S, {Singh} S~V, {Kakad} A~P, {Verheest} F, and {Bharuthram} R 2008
  {\em \npgeo\/} {\bf 15} 903--913

\bibitem{Lakhina2011}
{Lakhina} G~S, {Singh} S~V, and {Kakad} A~P 2011 {\em \adspr\/} {\bf 47}
  1558--1567

\bibitem{Gosling1989}
{Gosling} J~T, {Thomsen} M~F, {Bame} S~J, and {Russell} C~T 1989 {\em \jgr\/}
  {\bf 94} 10011--10025

\bibitem{Pierrard2010}
{Pierrard} V and {Lazar} M 2010 {\em \solphys\/} {\bf 267} 153--174

\bibitem{Vasyliunas1968}
{Vasyliunas} V~M 1968 {\em \jgr\/} {\bf 73} 2839--2884

\bibitem{Summers1991}
{Summers} D and {Thorne} R~M 1991 {\em \phflb\/} {\bf 3} 1835--1847

\bibitem{Baluku2008}
{Baluku} T~K and {Hellberg} M~A 2008 {\em \phpl\/} {\bf 15} 123705

\bibitem{Hellberg2009}
{Hellberg} M~A, {Mace} R~L, {Baluku} T~K, {Kourakis} I, and {Saini} N~S 2009
  {\em \phpl\/} {\bf 16} 094701

\bibitem{Sahu2010}
{Sahu} B 2010 {\em \phpl\/} {\bf 17} 122305

\bibitem{Younsi2010}
{Younsi} S and {Tribeche} M 2010 {\em \apss\/} {\bf 330} 295--300

\bibitem{Danehkar2011}
{Danehkar} A, {Saini} N~S, {Hellberg} M~A, and {Kourakis} I 2011 {\em \phpl\/}
  {\bf 18} 072902--072902

\bibitem{Devanandhan2011a}
{Devanandhan} S, {Singh} S~V, and {Lakhina} G~S 2011 {\em \physscr\/} {\bf 84}
  025507

\bibitem{Sultana2012}
{Sultana} S and {Kourakis} I 2012 {\em \epjd\/} {\bf 66} 100

\bibitem{Sultana2012b}
{Sultana} S, {Kourakis} I, and {Hellberg} M~A 2012 {\em \ppcf\/} {\bf 54}
  105016

\bibitem{Sahu2013}
{Sahu} B 2013 {\em \epl\/} {\bf 101} 55002

\bibitem{Han2014b}
{Han} J~N, {Duan} W~S, {Li} J~X, {He} Y~L, {Luo} J~H, {Nan} Y~G, {Han} Z~H, and
  {Dong} G~X 2014 {\em \phpl\/} {\bf 21} 012102

\bibitem{Han2014c}
{Han} J~N, {Duan} W~S, {Li} J~X, {Dong} G~X, and {Ge} S~H 2014 {\em \apss\/}
  {\bf 351} 525--531

\bibitem{Singh2016}
{Singh} S~V, {Devanandhan} S, {Lakhina} G~S, and {Bharuthram} R 2016 {\em
  \phpl\/} {\bf 23} 082310

\bibitem{Danehkar2011a}
{Danehkar} A, {Saini} N~S, {Hellberg} M~A, and {Kourakis} I 2011 {\em AIP Conf.
  Proc. (6th International Conference on the Physics of Dusty Plasmas)\/} vol
  1397 ed {Nosenko} V~Y, {Shukla} P~K, {Thoma} M~H, and {Thomas} H~M pp
  305--306 doi:10.1063/1.3659815

\bibitem{Danehkar2014}
{Danehkar} A, {Kourakis} I, and {Hellberg} M~A 2014 {\em IEEE 41st
  International Conference on Plasma Sciences (ICOPS), High-Power Particle
  Beams (BEAMS)\/} p 7012747 doi:10.1109/PLASMA.2014.7012747

\bibitem{Verheest2005}
{Verheest} F, {Cattaert} T, and {Hellberg} M~A 2005 {\em \ssr\/} {\bf 121}
  299--311

\bibitem{Verheest2007}
{Verheest} F, {Hellberg} M~A, and {Lakhina} G~S 2007 {\em \apsst\/} {\bf 3}
  15--20

\bibitem{Danehkar2009}
{Danehkar} A 2009 Master's thesis Queen's University Belfast, Northern Ireland,
  UK doi:10.5281/zenodo.47796

\bibitem{Saini2010}
{Saini} N~S and {Kourakis} I 2010 {\em \ppcf\/} {\bf 52} 075009

\bibitem{Saberian2013}
{Saberian} E, {Esfandyari-Kalejahi} A, {Rastkar-Ebrahimzadeh} A, and
  {Afsari-Ghazi} M 2013 {\em \phpl\/} {\bf 20} 032307

\bibitem{Jilani2014}
{Jilani} K, {Mirza} A~M, and {Khan} T~A 2014 {\em \apss\/} {\bf 349} 255--263

\bibitem{Danehkar2017}
{Danehkar} A 2017 {\em \phpl\/} {\bf 24} 102905

\bibitem{Schriver1989}
{Schriver} D and {Ashour-Abdalla} M 1989 {\em \grl\/} {\bf 16} 899--902

\bibitem{Verheest2009a}
{Verheest} F 2009 {\em \jpha\/} {\bf 42} 285501

\bibitem{Cattaert2005}
{Cattaert} T, {Verheest} F, and {Hellberg} M~A 2005 {\em \phpl\/} {\bf 12}
  042901

\end{thebibliography}
\end{document}